\documentclass[twocolumn,journal]{IEEEtran}
\usepackage[T1]{fontenc}
\usepackage[latin9]{inputenc}
\usepackage{color}
\usepackage{verbatim}
\usepackage{float}
\usepackage{bm}
\usepackage{amsmath}
\usepackage{amsthm}
\usepackage{amssymb}
\usepackage{graphicx}
\usepackage{subfigure}
\usepackage{caption}
\usepackage[unicode=true,
 bookmarks=true,bookmarksnumbered=true,bookmarksopen=true,bookmarksopenlevel=1,
 breaklinks=false,pdfborder={0 0 0},pdfborderstyle={},backref=false,colorlinks=false]
 {hyperref}
\hypersetup{pdftitle={Your Title},
 pdfauthor={Your Name},
 pdfpagelayout=OneColumn, pdfnewwindow=true, pdfstartview=XYZ, plainpages=false}
\usepackage{breakurl}

\makeatletter

\floatstyle{ruled}
\newfloat{algorithm}{tbp}{loa}
\providecommand{\algorithmname}{Algorithm}
\floatname{algorithm}{\protect\algorithmname}
\theoremstyle{plain}
\newtheorem{thm}{\protect\theoremname}
\theoremstyle{plain}
\newtheorem{lem}[thm]{\protect\lemmaname}

\makeatother

\providecommand{\lemmaname}{Lemma}
\providecommand{\theoremname}{Theorem}

\begin{document}

\title{Mixed-Timescale Beamforming and Power Splitting for Massive MIMO
Aided SWIPT IoT Network}

\author{{\normalsize{}Xihan Chen$^{1}$, Hei Victor Cheng$^{2}$, An Liu$^{1}$,
Kaiming Shen$^{2}$}\textit{\normalsize{},}{\normalsize{} and Min-Jian
Zhao$^{1}$}\thanks{This work was supported by the Science and Technology Program of Shenzhen,
China, under Grant JCYJ20170818113908577, and the National Natural
Science Foundation of China under Project No. 61571383. The work of
An Liu was supported by the China Recruitment Program of Global Young
Experts.

Xihan Chen, An Liu and Min-Jian Zhao are with the College of Information
Science and Electronic Engineering, Zhejiang University, Hangzhou
310027, China (e-mail: chenxihan@zju.edu.cn, anliu@zju.edu.cn, mjzhao@zju.edu.cn).

Hei Victor Cheng, and Kaiming Shen are with the Electrical and Computer
Engineering Department, University of Toronto, Toronto, ON M5S 3G4,
Canada (e-mail: hei.cheng@utoronto.ca, kshen@ece.utoronto.ca).}}
\maketitle
\begin{abstract}
Traditional simultaneous wireless information and power transfer (SWIPT)
with power splitting assumes perfect channel state information (CSI), which is difficult to obtain especially in the
massive multiple-input-multiple-output (MIMO) regime. In this letter,
we consider a mixed-timescale joint beamforming and power splitting
(MJBP) scheme to maximize general utility functions under a power
constraint in the downlink of a massive MIMO SWIPT IoT network. In
this scheme, the transmit digital beamformer is adapted to the imperfect
CSI, while the receive power splitters are adapted to the long-term
channel statistics only due to the consideration of hardware limit
and signaling overhead. The formulated optimization problem is solved
using a mixed-timescale online stochastic successive convex approximation
(MO-SSCA) algorithm. Simulation results reveal significant gain over
the baselines.
\end{abstract}


\IEEEpeerreviewmaketitle{}

\section{Introduction}

The Internet of Things (IoT) \cite{Swan_survey12_IoT} is a revolutionary
communication paradigm to provide massive connectivity for the next-generation
wireless cellular networks. The limited battery life of devices poses
a significant challenge for designing green and sustainable IoT. One
promising solution is to leverage the simultaneous wireless information
and power transfer (SWIPT) with radio frequency to prolong the IoT
network, due to its ability to provide cost-effective and perpetual
power source \cite{Ruizhang_2013twc_SpatialSWIPT}. This requires
receiver circuits to decode information and harvest energy from the
same received signal independently and simultaneously, which renders
SWIPT impractical.

To overcome these limitations, the telecommunication industry is increasingly
turning towards power splitting (PS), a receiver architecture that
divides the received signal into two streams of different power for
decoding information and harvesting energy. Based on the PS architecture,
\cite{QJS_2014twc_JBPS} considers a multiuser joint beamforming and
power splitting design problem under QoS constraints and proposes
a semidefinite relaxation-based algorithm. To further reduce the computational
complexity, an second-order cone programming relaxation method is
proposed in \cite{QJS_2014tsp_SOCP}. Recently, \cite{Dong_2018globecom_ZFSWIPT}
combines the SWIPT and massive multiple-input-multiple-output (MIMO)
to further improve the spectral and energy efficiency of IoT networks.
The aforementioned works focus on optimizing the weighted sum of objective
function under perfect CSI, which is difficult to obtain in the massive
MIMO regime due to the large number of antennas and the limited pilot
sequences\textcolor{black}{{} \cite{Caire_2018twc_MIMObound}}. In such
scenarios, it is more reasonable to consider a mixed-timescale optimization
of the long-term performance of the network, which only requires imperfect
CSI plus the knowledge of channel statistics \cite{Anliu_tsp19_THCF}.
To the best of our knowledge, this is first work on mixed-timescale
optimization for massive MIMO aided SWIPT IoT network.%
\begin{comment}
To reduce the CSI signaling overhead, work \cite{Liu_TSP13_CacheIFN}
proposes a mixed-timescale optimization scheme for cached MIMO network,
where the cache content placement is updated at a slower timescale
than the other control variables. However, there is no work on mixed-timescale
optimization for massive MIMO aided SWIPT system.
\end{comment}

Contribution of this letter includes the algorithm design for mixed-timescale
joint beamforming and power splitting (MJBP) scheme for the downlink
transmission of massive MIMO aided SWIPT IoT network, to maximize
a general network utility. Specifically, the digital beamformer is
adapted to the imperfect CSI, while the power spiltters are adapted
to the long-term channel statistics due to the consideration of hardware
limit and signaling overhead. We propose a \textit{mixed-timescale
online stochastic successive convex approximation} (MO-SSCA) algorithm
to solve this joint optimization problem. Simulations verify the advantages
of the proposed MJBP scheme over the baselines.

\begin{figure}[tbh]
\begin{centering}
\includegraphics[width=1\columnwidth,height=2.8cm]{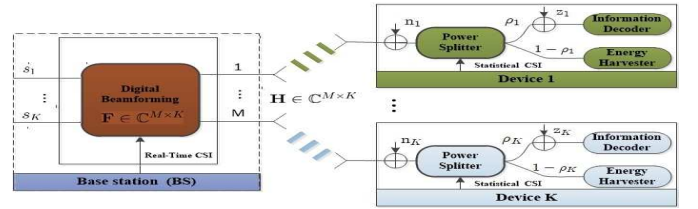}
\par\end{centering}
\caption{\label{fig:MBPS}Architecture of the proposed MJBP scheme.}
\end{figure}

\section{System Model and Problem Formulation}

Consider the downlink of a massive MIMO aided SWIPT IoT network, where
the base station (BS) is equipped with $M$ antennas to simultaneously
serve $K$ single-antenna IoT devices. As illustrated in Fig. \ref{fig:MBPS},
the BS employs digital beamformer $\mathbf{F}\triangleq[\bm{f}_{1},\cdots,\bm{f}_{K}]\in\mathbb{C}^{M\times K}$
to spatially multiplex devices and manage the multi-device interference,
while device $k$ applies the power splitter $\rho_{k}$ ($0\leq\rho_{k}\leq1$)
to coordinate information decoding and energy harvesting from the
received signal. With MJBP, both the digital beamformer and the power
splitters are optimized at the BS. Furthermore, the digital beamformer
is adapted to instantaneous CSI. For the power splitter implemented
at each device, it is adapted to long-term channel statistics due
to following reasons: 1) the hardware capability
of the IoT device is limited, and thus the power splitter cannot be
changed frequently due to hardware limitations \cite{Swan_survey12_IoT};
2) such design can reduce the signaling overhead of sending $\rho_{k},\forall k,$
to the corresponding device, especially when the number of devices
is large.

We consider flat fading channels with block fading model, but the
proposed algorithm can be easily modified to cover the frequency selective
channels. The channel $\mathbf{H}\triangleq[\bm{h}_{1},\cdots,\bm{h}_{K}]\in\mathbb{C}^{M\times K}$
is assumed to be constant within each block of length
$T$. In this case, the received signal splitted to the information
decoder (ID) of device $k$ is given by $y_{k}^{I}=\sqrt{\rho_{k}}(\bm{h}_{k}^{H}\sum_{m=1}^{K}\bm{f}_{m}s_{m}+n_{k})+z_{k},$
where $s_{m}\sim\mathcal{CN}(0,1)$ is the data symbol for device
$m$, $n_{k}\sim\mathcal{CN}(0,\sigma_{k}^{2})$ is the additive noise
(AN) at the PS of device $k$, and $z_{k}\sim\mathcal{CN}(0,\delta_{k}^{2})$
is the AN introduced by the ID at device $k$. Meanwhile, the received
signal splitted to the energy harvester (EH) is given by $y_{k}^{E}=\sqrt{1-\rho_{k}}(\bm{h}_{k}^{H}\sum_{m=1}^{K}\bm{f}_{m}s_{m}+n_{k}).$

An achievable ergodic rate \cite{Caire_2018twc_MIMObound}
at device $k$ is given by
\begin{align}
\hat{r}_{k}^{\circ}(\rho_{k},\mathbf{F}) & =\mathbb{E}_{\mathbf{H}}[\log_{2}(1+\textrm{\ensuremath{\Gamma}}_{k}(\rho_{k},\mathbf{F},\mathbf{H}))]\nonumber \\
 & -\frac{1}{T}\sum_{m=1}^{K}\log_{2}(1+\frac{T}{\rho_{k}\sigma_{k}^{2}+\delta_{k}}\mathrm{Var}(\bm{h}_{k}^{H}\bm{f}_{m}),\label{eq:ergodiRate}
\end{align}
where $\textrm{\ensuremath{\Gamma}}_{k}(\rho_{k},\mathbf{F},\mathbf{H})$
is the SINR of device $k$ with
\[
\textrm{\ensuremath{\Gamma}}_{k}(\rho_{k},\mathbf{F},\mathbf{H})=\frac{\rho_{k}|\bm{h}_{k}^{H}\bm{f}_{k}|^{2}}{\rho_{k}(\sum_{m\neq k}^{K}|\bm{h}_{k}^{H}\bm{f}_{m}|^{2}+\sigma_{k}^{2})+\delta_{k}^{2}}.
\]

\textcolor{black}{In practice, perfect CSI is challenging to obtain
due to device mobility, processing latency and other limitations.
Thus, we model the channel imperfection as $\bm{h}_{k}=\hat{\bm{h}}_{k}+\bm{\phi}_{k},$
where $\hat{\bm{h}}_{k}$ is the estimated channel from from BS to
device $k$, $\bm{\phi}_{k}\sim\mathcal{CN}(0,\omega_{k}^{2}\mathbf{I}_{M})$
is the channe}l error independent of $\hat{\bm{h}}_{k}$, and $\omega_{k}^{2}$
is the variance of the channel error. Consequently,
the achievable rate is obtained by replacing $\bm{h}_{k}$ in (\ref{eq:ergodiRate})
with $\hat{\bm{h}}_{k}+\bm{\phi}_{k}$. \textcolor{black}{For convenience,
we let $\mathbf{\hat{H}}\triangleq[\bm{\hat{h}}_{1},\cdots,\hat{\bm{h}}_{K}]$,
and $\bm{\phi}\triangleq[\bm{\phi}_{1}^{T},\cdots,\bm{\phi}_{K}^{T}]^{T}$.
Further, we define $\bm{\Theta}\triangleq\{\mathbf{F}(\mathbf{\hat{H}})\in\varLambda,\forall\mathbf{\hat{H}}\}$
as the collection of short-term optimization variables for all possible
estimated channel states $\mathbf{\hat{H}}$, where $\varLambda\triangleq\{\mathbf{F}|\textrm{Tr}(\mathbf{F}\mathbf{F}^{H})\leq P_{\textrm{max}}\}$
is the feasible set of $\mathbf{F}$. }

\emph{Proposition 1} : The ergodic rate at device
is $k$ bounded as $\hat{r}_{k}^{\circ}(\rho_{k},\mathbf{\Theta})\leq\overline{r}_{k}(\rho_{k},\mathbf{\Theta})\triangleq\mathbb{E}_{\mathbf{\hat{H}},\bm{\phi}_{k}}[\log_{2}(1+\textrm{\ensuremath{\Gamma}}_{k}(\rho_{k},\mathbf{F},\hat{\bm{h}}_{k},\bm{\phi}_{k}))]$,
and $\hat{r}_{k}^{\circ}(\rho_{k},\mathbf{\Theta})\geq$
\[
\overline{r}_{k}(\rho_{k},\mathbf{\Theta})-\frac{1}{T}\sum_{m=1}^{K}\log_{2}(1+\frac{TP_{\textrm{max}}}{\delta_{k}^{2}}\mathbb{E}_{\mathbf{\hat{H}},\bm{\phi}_{k}}[||\hat{\bm{h}}_{k}+\bm{\phi}_{k}||^{2}]).
\]

Here the lower bound follows from the properties of variance and the
Cauchy--Schwarz inequality. From Proposition 1, optimizing the lower and upper
bound provide the same optimal solution. Moreover, as verified in Fig. 2,
we find that both bounds are tight. Therefore, we optimize
the lower (upper) bound of the ergodic rate at each device as it is more tractable for optimization.

\begin{figure}[tbh]
\begin{centering}
\includegraphics[height=3cm]{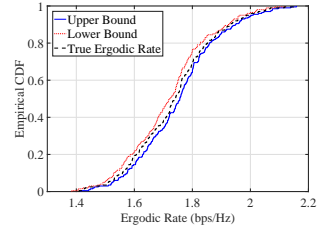}
\par\end{centering}
\caption{Cumulative distribution function (CDF) of the lower
and upper bound of ergodic rate at each device. The detailed setup
is given in Section \ref{sec:Simulation-Results-and}.}
\end{figure}

\textcolor{black}{The average harvested power conditioned on imperfect
CSI $\hat{\bm{h}}_{k}$ of device $k$ follows a non-linear function
\cite{Ng_CL15_nonlinear} and can be expressed as $\overline{e}_{k}^{\circ}(\rho_{k},\mathbf{F}|\hat{\bm{h}}_{k})=\mathbb{E}_{\bm{\phi}{}_{k}}[\hat{e}_{k}^{\circ}(\rho_{k},\mathbf{F}|\hat{\bm{h}}_{k},\bm{\phi}_{k})],$}\textcolor{black}{{}
where}\textcolor{black}{
\[
\hat{e}_{k}^{\circ}(\rho_{k},\mathbf{F}|\hat{\bm{h}}_{k},\bm{\phi}_{k})=(\Psi_{k}-S_{k}\Omega_{k})/(1-\Omega_{k}),
\]
where $S_{k}$ is a constant denoting the maximum harvested power
at the $k$th device, $\Psi_{k}\triangleq\frac{S_{k}}{1+\mathrm{exp}(-a_{k}(P_{k}-b_{k}))}$,
$\Omega_{k}\triangleq\frac{1}{1+\mathrm{exp}(a_{k}b_{k})}$, and parameter
$a_{k}$ and $b_{k}$ are constants related to the circuit specifications,
and $P_{k}\triangleq(1-\rho_{k})(\sum_{m=1}^{K}|(\hat{\bm{h}}_{k}+\bm{\phi}_{k})^{H}\bm{f}_{m}|^{2}+\sigma_{k}^{2})$
is the input RF power for the $k$th device.}\textcolor{black}{{} Then,
the average harvested power of user $k$ is defined as }\textcolor{black}{$\overline{e}_{k}(\rho_{k},\mathbf{\Theta})=\mathbb{E}_{\mathbf{\hat{H}}}[\overline{e}_{k}^{\circ}(\rho_{k},\mathbf{F}\left(\mathbf{\hat{H}}\right)|\hat{\bm{h}}_{k})]$.}

\textcolor{black}{We are interested in a mixed-timescale joint optimization
of digital beamformer and power splitter to balance the average ergodic
rate and the average harvested power. This can be formulated as the
following network utility maximization problem:
\begin{equation}
\mathcal{P}:\max_{\boldsymbol{\rho}\in\Phi,\Theta}\:\sum_{k=1}^{K}g\left(\overline{\eta}_{k}\left(\rho_{k},\bm{\Theta}\right)\right),\label{eq:originalP}
\end{equation}
}where $\overline{\eta}_{k}(\rho_{k},\mathbf{\Theta})\triangleq\overline{r}_{k}(\rho_{k},\mathbf{\Theta})+\gamma_{k}\overline{e}_{k}(\rho_{k},\mathbf{\Theta})$
with the corresponding weight $\gamma_{k}$ is a weighted sum of the
average \textcolor{black}{ergodic} rate and the harvested power, $\Phi\triangleq\{\bm{\rho}=[\rho_{1},\cdots,\rho_{K}]^{T}|\rho_{k}\in(0,1],\forall k\}$
is the feasible set of power splitters. The utility function $g\left(\overline{\eta}_{k}\right)$
is a continuously differentiable and concave function of $\overline{\eta}_{k}$.
Moreover, $g\left(\overline{\eta}_{k}\right)$ is non-decreasing w.r.t.
$\overline{\eta}_{k},$ and its derivative $\nabla_{\overline{\eta}_{k}}g\left(\overline{\eta}_{k}\right)$
is Lipschitz continuous.

\section{Online Optimization Algorithm}

In this section, we propose a MO-SSCA algorithm to solve the mixed-timescale
stochastic non-convex optimization problem $\mathcal{P}$, and summarize
it in Algorithm 1. In MO-SSCA, we focus on a coherence time of channel
statistics, where the time is divided into $T_{f}$ frames and each
frame consists of $T_{s}$ time slots. At beginning, the BS initializes
the MO-SSCA algorithm with power splitter $\bm{\rho}^{0}$ and a weight
vector $\bm{v}^{0}$. In subsequent, $\bm{\rho}$ and $\bm{v}$ are
updated once at the end of each frame. Then we elaborate the implementation
details of the iteration of the MO-SSCA algorithm at the $t$-th frame.

\begin{algorithm}[tbh]
\caption{\label{alg1}MO-SSCA Algorithm}

\textbf{\small{}Input: }{\small{}$\left\{ \alpha^{t}\right\} $, $\left\{ \beta^{t}\right\} $.}{\small\par}

\textbf{\small{}Initialize:}{\small{} $\boldsymbol{\rho}^{0}\in\Phi$;
$\boldsymbol{v}^{0}=\left[1,...,1\right]^{T}$, $t=0$.}{\small\par}

\textbf{\small{}Step 1 }{\small{}(}\textbf{\small{}Short-term optimization
at each time slot}{\small{} $i\in\mathcal{T}_{t}$):}{\small\par}

\textbf{\small{}\,\,\,\,\,\,\,\,\,\,}\textcolor{black}{\small{}Apply
the short-term algorithm with input $\boldsymbol{v}^{t},\boldsymbol{\rho}^{t}$
and $\boldsymbol{\hat{H}}(i)$, to obtain the short-term variable
$\mathbf{F}^{t}(i)$, as elaborated in Section \ref{subsec:short-term algorithm}.}{\small\par}

\textbf{\textcolor{black}{\small{}Step 2 (Long-term optimization at
the end of frame $t$)}}\textcolor{black}{\small{}: }{\small\par}

\textcolor{black}{\small{}\,\,\,\,\,\,\,\,\,\,}\textbf{\textcolor{black}{\small{}2a:}}\textcolor{black}{\small{}
}\textbf{\textcolor{black}{\small{}Obtain}}\textcolor{black}{\small{}
the estimated channel sample $\bm{\hat{h}}_{k}^{t}\triangleq\bm{\hat{h}}_{k}(tT_{s}+1),\forall k$.}{\small\par}

\textcolor{black}{\small{}\,\,\,\,\,\,\,\,\,\,}\textbf{\textcolor{black}{\small{}2b:
Update}}\textcolor{black}{\small{} the surrogate function $\overline{g}^{t}(\rho_{k})$
according to (\ref{upsurrgate}). }{\small\par}

\textcolor{black}{\small{}\,\,\,\,\,\,\,\,\,\,}\textbf{\textcolor{black}{\small{}2c:
Calculate}}\textcolor{black}{\small{} $\bar{\boldsymbol{v}}^{t}\triangleq\nabla_{\overline{\bm{\eta}}}g\left(\hat{\bm{\eta}}^{t}\right)$
and update $\boldsymbol{v}^{t+1}$ according to (\ref{eq:updatemu}).}{\small\par}

\textcolor{black}{\small{}\,\,\,\,\,\,\,\,\,\,}\textbf{\textcolor{black}{\small{}2d:
Solve}}\textcolor{black}{\small{} (\ref{eq:Pitert}) to obtain $\bar{\boldsymbol{\rho}_{k}}^{t}$
and update $\boldsymbol{\rho}_{k}^{t+1}$ according to (\ref{eq:update-rho}).}{\small\par}

\textbf{\textcolor{black}{\small{}Let}}\textcolor{black}{\small{}
$t=t+1$ and return to }\textbf{\textcolor{black}{\small{}Step 1}}\textcolor{black}{\small{}.}{\small\par}
\end{algorithm}

\subsection{Short-term FP-BCD Algorithm\label{subsec:short-term algorithm}}

At time slot $i\in\mathcal{T}_{t}\triangleq[tT_{s}+1,(t+1)T_{s}]$
within the $t$-th frame, BS obtains the estimated channel $\mathbf{\hat{H}}(i)$
by uplink channel training. Based upon the current $\bm{\rho}^{t}$,
$\bm{v}^{t}$, and $\mathbf{\hat{H}}(i)$, we can obtain digital beamforming
by maximizing the a weighted sum of the average data rate and the
average harvested power conditioned on imperfect CSI, which can be
formulated as
\[
\mathcal{P}_{2}\left(\bm{\rho},\boldsymbol{v},\mathbf{\hat{H}}\right):\max_{\mathbf{\mathbf{F}}\in\varLambda}\:\sum_{k=1}^{K}v_{k}\overline{\eta}_{k}^{\circ}(\rho_{k},\mathbf{F}|\bm{\hat{h}}_{k}),
\]
where $\overline{\eta}_{k}^{\circ}(\rho_{k},\mathbf{F}|\bm{\hat{h}}_{k})\triangleq\overline{r}_{k}^{\circ}(\rho_{k},\mathbf{F}|\bm{\hat{h}}_{k})+\text{\ensuremath{\gamma_{k}}}\overline{e}_{k}^{\circ}(\rho_{k},\mathbf{F}|\bm{\hat{h}}_{k})$
\textcolor{black}{with $\overline{r}_{k}^{\circ}(\rho_{k},\mathbf{F}|\bm{\hat{h}}_{k})=\mathbb{E}_{\bm{\phi}}[\hat{r}_{k}^{\circ}(\rho_{k},\mathbf{F}|\hat{\bm{h}}_{k},\bm{\phi}_{k})]$}
\textcolor{black}{and $\hat{r}_{k}^{\circ}(\rho_{k},\mathbf{F}|\hat{\bm{h}}_{k},\bm{\phi}_{k})=\log_{2}(1+\textrm{\ensuremath{\Gamma}}_{k}(\rho_{k},\mathbf{F},\hat{\bm{h}}_{k},\bm{\phi}_{k}))$},
and $\mathcal{P}_{2}\left(\bm{\rho}^{t},\bm{v}^{t},\mathbf{\hat{H}}(i)\right)$
is solved at time slot $i\in\mathcal{T}_{t}$.

Since the objective function $\overline{\eta}_{k}^{\circ}(\rho_{k},\mathbf{F}|\bm{\hat{h}}_{k})$
contains expectation operators, it d\textcolor{black}{oes not have
a closed-form expression. To address the challenge, we resort to the
Sample Average Approximation (SAA) method \cite{Shapiro_SIAM09_SAA}.
Specifically, a total of $N$ samples are generated for $\bm{\phi}_{k}$
independently drawn from the distribution $\mathcal{CN}(0,\omega_{k}^{2}\mathbf{I}_{M})$,
and the $n$-th sample of $\bm{\phi}_{k}$ is defined as $\bm{\phi}_{k}^{n}.$
In this case, the SAA version of $\mathcal{P}_{2}\left(\bm{\rho},\boldsymbol{v},\mathbf{\hat{H}}\right)$
is formulated as $\mathcal{P}_{3}\left(\bm{\rho},\boldsymbol{v},\mathbf{\hat{H}}\right):\max_{\mathbf{F}\in\varLambda}\:\frac{1}{N}\sum_{k=1}^{K}\sum_{n=1}^{N}v_{k}\hat{\eta}_{k}^{\circ}(\rho_{k},\mathbf{F}|\bm{\hat{h}}_{k},\bm{\phi}_{k}^{n}),$
where $\hat{\eta}_{k}^{\circ}(\rho_{k},\mathbf{F}|\bm{\hat{h}}_{k},\bm{\phi}_{k}^{n})=\hat{r}_{k}^{\circ}(\rho_{k},\mathbf{F}|\hat{\bm{h}}_{k},\bm{\phi}_{k}^{n})+\gamma_{k}\hat{e}_{k}^{\circ}(\rho_{k},\mathbf{F}|\hat{\bm{h}}_{k},\bm{\phi}_{k}^{n})$.}

However, solving problem $\mathcal{P}_{3}\left(\bm{\rho},\boldsymbol{v},\mathbf{\hat{H}}\right)$
is still challenging due to the nonlinear fractional term in $\hat{r}_{k}$
and coupling in the power constraint. To this end, we apply the Lagrangian
dual transform method \cite{Kaiming_TSP18_FP} to recast problem $\mathcal{P}_{3}\left(\bm{\rho},\boldsymbol{v},\mathbf{\hat{H}}\right)$
into a more tractable yet equivalent form, using the following lemma.
\begin{lem}
\label{lem:transformation-F} The optimal digital beamforming $\mathbf{\mathbf{F}}^{\ast}$
solves the problem in (\ref{lem:transformation-F}) if and only if
it solves
\begin{equation}
\max_{\mathbf{\mathbf{\mathbf{F}}}\in\varLambda}\:\frac{1}{N}\sum_{n=1}^{N}\sum_{k=1}^{K}f\left(\mathbf{F},q_{k}^{n},\bm{\phi}_{k}^{n}\right)\label{eq:trans-short-sub}
\end{equation}
where\textcolor{black}{{} $f\left(\mathbf{F},q_{k}^{n},\bm{\phi}_{k}^{n}\right)\triangleq v_{k}(\log_{2}(1+q_{k}^{n})-q_{k}^{n}+(\rho_{k}(\Gamma_{k}^{n}+\sigma_{k}^{2})+\delta_{k}^{2})^{-1}(1+q_{k}^{n})\rho_{k}|\tilde{\bm{h}}_{k,n}^{H}\bm{f}_{k}|^{2}+\gamma_{k}\hat{e}_{k}^{\circ}(\mathbf{F},q_{k}^{n},\bm{\phi}_{k}^{n},w_{k}^{n})),$}
$\Gamma_{k}^{n}\triangleq\sum_{m=1}^{K}|\tilde{\bm{h}}_{k,n}^{H}\bm{f}_{m}|^{2}$
with $\tilde{\bm{h}}_{k,n}\triangleq\bm{\phi}_{k}^{n}+\bm{\hat{h}}_{k}$,
and $q_{k}^{n}\triangleq\frac{\rho_{k}|\tilde{\bm{h}}_{k,n}^{H}\bm{f}_{k}|^{2}}{\rho_{k}\left(\sum_{m\neq k}^{K}|\tilde{\bm{h}}_{k,n}^{H}\bm{f}_{m}|^{2}+\sigma_{k}^{2}\right)+\delta_{k}^{2}}$
is the optimal auxiliary variable introduced for each ratio term.
\end{lem}
In subsequent, we use the complex quadratic transformation \cite{Kaiming_TSP18_FP}
to equivalently recast problem (\ref{eq:trans-short-sub}) as \textcolor{black}{
\begin{equation}
\max_{\mathbf{\mathbf{\mathbf{F}}}\in\varLambda,\bm{w},\bm{q}}\:\frac{1}{N}\sum_{n=1}^{N}\sum_{k=1}^{K}\hat{r}_{k}\left(\mathbf{F},q_{k}^{n},\bm{\phi}_{k}^{n},w_{k}^{n}\right)+\hat{e}_{k}^{\circ}\left(\mathbf{F},q_{k}^{n},\bm{\phi}_{k}^{n},w_{k}^{n}\right)\label{eq:final-sub}
\end{equation}
}where
\begin{align*}
 & \hat{r}_{k}\left(\mathbf{F},q_{k}^{n},\bm{\phi}_{k}^{n},w_{k}^{n}\right)\triangleq\sqrt{v_{k}\text{\ensuremath{\rho_{k}}}(1+q_{k}^{n})}\mathrm{Re}\left\{ \bm{f}_{k}^{H}\tilde{\bm{h}}_{k,n}\left(w_{k}^{n}\right)^{H}\right\} \\
 & +\left(w_{k}^{n}\right)^{H}w_{k}^{n}\left(\rho_{k}(\Gamma_{k}^{n}+\sigma_{k}^{2})+\delta_{k}^{2}\right)-q_{k}^{n}+\log_{2}(1+q_{k}^{n})\text{.}
\end{align*}
$\bm{q}=[\bm{q}_{1}^{T},\cdots,\bm{q}_{K}^{T}]^{T}$ with $\bm{q}_{k}=[q_{k}^{1},\cdots,q_{k}^{N}]^{T}$,
and $\bm{w}=[\bm{w}_{1}^{T},\cdots,\bm{w}_{K}^{T}]^{T}$ with $\bm{w}_{k}=[w_{k}^{1},\cdots,w_{k}^{N}]^{T}$
is the auxiliary variable vector. Observing that the constraints are
separable with respect to the three blocks of variables, i.e., $\bm{q},$
$\bm{w}$, and $\mathbf{F}$, we shall focus on designing a \textcolor{black}{fractional
programming block coordinate descent (FP-BCD)} algorithm to find a
stationary point of problem (\ref{eq:final-sub}), and summarize it
in Algorithm 2. For problem (\ref{eq:final-sub}), this amounts to
the following steps:

\begin{algorithm}[tbh]
\caption{Short-term FP-BCD Algorithm for problem (\ref{eq:final-sub})}

\textbf{\small{}Input:}{\small{} $\boldsymbol{v},\boldsymbol{\rho}$,
$\mathbf{\hat{H}}$, the sample number $N$.}{\small\par}

\textbf{\small{}Initialization:}{\small{} Initialize $\mathbf{F}$
to feasible values.}{\small\par}

\textbf{\textcolor{black}{Repeat}}

\textbf{\small{}\,\,\,\,\,\,\,\,\,\,\,\,\,Step 1 : Update}{\small{}
$\boldsymbol{q}$ according to (\ref{eq:update-q}).}{\small\par}

\textbf{\small{}\,\,\,\,\,\,\,\,\,\,\,\,\,Step 2 : Update}{\small{}
$\boldsymbol{w}$ according to (\ref{eq:update-W}).}{\small\par}

\textbf{\small{}\,\,\,\,\,\,\,\,\,\,\,\,\,Step 3 : Update}{\small{}
$\mathbf{F}$ by solvi}\textcolor{black}{\small{}ng problem (}\textcolor{black}{\ref{eq:final_F-1}}\textcolor{black}{\small{})
using CVX.}{\small\par}

\textbf{\textcolor{black}{until }}\textcolor{black}{the value of objective
function in (\ref{eq:final-sub}) converges}
\end{algorithm}

\subsubsection{Optimization of $\boldsymbol{q}$}

The optimal $\bm{q}^{\ast}$ is given by
\begin{equation}
\left(q_{k}^{n}\right)^{\ast}=\frac{\rho_{k}|\tilde{\bm{h}}_{k,n}^{H}\bm{f}_{k}|^{2}}{\rho_{k}(\Gamma_{k}^{n}-|\tilde{\bm{h}}_{k,n}^{H}\bm{f}_{k}|^{2}+\sigma_{k}^{2})+\delta_{k}^{2}}.\label{eq:update-q}
\end{equation}

\subsubsection{Optimization of $\boldsymbol{w}$}

By applying the first-order optimal condition, the optimal $\bm{w}^{\ast}$
admits a closed-form solution as:
\begin{equation}
\left(w_{k}^{n}\right)^{\ast}=\left(\rho_{k}(\Gamma_{k}^{n}+\sigma_{k}^{2})+\delta_{k}^{2}\right)^{-1}\!\!\!\!\!\sqrt{\rho_{k}v_{k}(1+q_{k}^{n})}\tilde{\bm{h}}_{k,n}^{H}\bm{f}_{k}.\label{eq:update-W}
\end{equation}

\subsubsection{Optimization of $\mathbf{F}$}

\textcolor{black}{The subproblem w.r.t. $\mathbf{F}$ is nonconvex
due to the involvement of the non-linear energy harvesting model.
To overcome this difficulty, we first transform it into a more tractable
yet equivalent form by the introduction of new auxiliary variables
$0\leq\varsigma_{k}^{n}\leq\hat{e}_{k}^{\circ}\left(\mathbf{F},q_{k}^{n},\bm{\phi}_{k}^{n},w_{k}^{n}\right)$
and some manipulations, which can be expressed as
\begin{align}
 & \mathcal{P}_{4}:\max_{\mathbf{\mathbf{\mathbf{F}}}\in\varLambda,\alpha_{k}^{n}\geq0}\frac{1}{N}\sum_{n=1}^{N}\sum_{k=1}^{K}\hat{r}_{k}\left(\mathbf{F},q_{k}^{n},\bm{\phi}_{k}^{n},w_{k}^{n}\right)+\varsigma_{k}^{n}\label{eq:P4}\\
 & \mathrm{s.t.}\ln(1/(\alpha_{k}^{n}+\Omega_{k}^{n})\!-\!1)\!+\!d_{k}\sum_{m=1}^{K}|(\hat{\bm{h}}_{k}+\bm{\phi}_{k}^{n})^{H}\bm{f}_{m}|^{2}\!+\!c_{k}\!\geq\!0,\nonumber
\end{align}
where $d_{k}\triangleq a_{k}(1-\rho_{k})$, and $c_{k}\triangleq a_{k}\sigma_{k}^{2}(1-\rho_{k})-a_{k}b_{k}$.
Note that the constraint in problem $\mathcal{P}_{4}$ is nonconvex.
Thus, we apply the the majorization minimization (MM) method \cite{Jacobson_TIP07_MM}
to approximate this nonconvex constraint using its first-order Taylor
expansion as
\begin{align}
 & \max_{\mathbf{\mathbf{\mathbf{F}}}\in\varLambda,\alpha_{k}^{n}\geq0}\frac{1}{N}\sum_{n=1}^{N}\sum_{k=1}^{K}\hat{r}_{k}\left(\mathbf{F},q_{k}^{n},\bm{\phi}_{k}^{n},w_{k}^{n}\right)+\varsigma_{k}^{n}\label{eq:final_F-1}\\
 & \mathrm{s.t.}\:d_{k}\sum_{m=1}^{K}(\hat{\bm{h}}_{k}+\bm{\phi}_{k}^{n})^{H}(\bm{\tilde{f}}_{m}\bm{\tilde{f}}_{m}^{H}\!+\!\bm{\tilde{f}}_{m}\bm{\overline{f}}_{m}^{H}\!+\!\bm{\overline{f}}_{m}\bm{\tilde{f}}_{m}^{H})(\hat{\bm{h}}_{k}+\bm{\phi}_{k}^{n})\nonumber \\
 & +\!\ln(1/(\tilde{\varsigma}_{k}^{n}+\Omega_{k}^{n})\!-\!1)\!+\!\frac{\varsigma_{k}^{n}-\tilde{\varsigma}_{k}^{n}}{(\tilde{\varsigma}_{k}^{n}+\Omega_{k}^{n}-1)(\tilde{\varsigma}_{k}^{n}+\Omega_{k}^{n})}\!+\!c_{k}\!\geq0,\nonumber
\end{align}
where $\tilde{\varsigma}_{k}^{n}$ and $\tilde{\bm{f}_{m}}$ represents
the last iteration of $\varsigma_{k}^{n}$ and $\bm{f}_{m}$, and
$\bm{\overline{f}}_{m}\triangleq$$\bm{f}_{m}-\tilde{\bm{f}}_{m}$.
Note that problem (\ref{eq:final_F-1}) is convex, which can be efficiently
solved by the CVX toolbox \cite{CVX_MATLAB}.}

\subsection{Long-term Optimization}

Before the end of $t$-th frame, device $k$ obtains a full channel
sample $\bm{\hat{h}}_{k}^{t}\triangleq\bm{\hat{h}}_{k}(tT_{s}+1)$
and channel error sample $\bm{\phi}_{k}^{t}$. Based on $\bm{\hat{h}}_{k}^{t}$,
$\bm{\phi}_{k}^{t}$ and $\mathbf{F}^{t}(i),\forall i\in\mathcal{T}_{t}$,
we preserve the partial concavity of the original function and add
the proximal regularization, to construct the concave surrogate function
$\overline{g}^{t}(\rho_{k})$, resulting in the following

\begin{equation}
\overline{g}^{t}(\rho_{k})=g(\tilde{\eta}_{k}^{t})+\left(u_{k}^{t}\right)^{T}(\rho_{k}-\rho_{k}^{t})-\tau|\rho_{k}-\rho_{k}^{t}|^{2},\label{upsurrgate}
\end{equation}
where $\tau>0$ is a postive constant; the recursive approximation
of the weighted sum of the data rate and the harvested power $\tilde{\eta}_{k}$
is given by
\[
\tilde{\eta}_{k}^{t}=(1-\alpha_{t})\tilde{\eta}_{k}^{t-1}+\frac{\alpha_{t}}{N}\sum_{n=1}^{N}\sum_{i\in\mathcal{T}_{t}}\frac{\hat{\eta}_{k}^{\circ}(\rho_{k}^{t},\mathbf{F}^{t}(i)|\bm{\hat{h}}_{k}^{i},\bm{\phi}_{k}^{n}(i))}{|\mathcal{T}_{t}|},
\]
with $\tilde{\eta}_{k}^{-1}=0$, and $\alpha_{t}\in\left(0,1\right]$
is a step-sizes sequence to be properly chosen; the recursive approximation
of the partial derivative $\nabla_{\rho_{k}}g\left(\tilde{\eta}_{k}\right)$
is given by
\[
u_{k}^{t}=(1-\alpha_{t})u_{k}^{t-1}+\alpha_{t}J_{\rho_{k}}\left(\rho_{k}^{t},\mathbf{F}^{t}(i)|\bm{\hat{h}}_{k}^{t},\bm{\phi}_{k}^{t}\right)\nabla_{\overline{\eta}_{k}}g\left(\tilde{\eta}_{k}^{t}\right),
\]
with $u_{k}^{-1}=0$, $J_{\rho_{k}}\left(\rho_{k}^{t},\mathbf{F}^{t}(i)|\bm{\hat{h}}_{k}^{t},\bm{\phi}_{k}^{t}\right)$
is the gradient of $\hat{\eta}_{k}^{\circ}\left(\rho_{k}^{t},\mathbf{F}^{t}(i)|\bm{\hat{h}}_{k}^{t},\bm{\phi}_{k}^{t}\right)$
w.r.t. $\rho_{k}$ at $\rho_{k}=\rho_{k}^{t}$ and $\mathbf{F}=\mathbf{F}^{t}(i)$.
Moreover, the weight vector $\boldsymbol{v}$ is updated as

\begin{equation}
v_{k}^{t+1}=\left(1-\beta_{t}\right)v_{k}^{t}+\beta_{t}\bar{v}_{k}^{t},\label{eq:updatemu}
\end{equation}
with $v_{k}^{t}=\nabla_{\overline{\eta}}g_{k}\left(\tilde{\eta}_{k}^{t}\right)$,
where $\beta_{t}\in\left(0,1\right]$ is a step-sizes sequence satisfying
$\sum_{t}\beta_{t}=\infty$, $\sum_{t}\left(\beta_{t}\right)^{2}<\infty$.
Moreover, the optimal power splitting ratio for device $k$ can be
obtained by solving the following quadratic optimization problem,
i.e.,

\begin{align}
\underset{\rho_{k}\in\Phi}{\text{max}}\: & \bar{g}^{t}\left(\rho_{k}\right).\label{eq:Pitert}
\end{align}
By applying the first-order optimality condition, it yields the\textcolor{black}{{}
closed-form solution $\bar{\rho}_{k}^{t}=\mathbb{P}_{\Phi}\left[\rho_{k}^{t}+\frac{u_{k}^{t}}{2\tau}\right],$
where $\mathbb{P}_{\Phi}\left[\cdot\right]$ denotes the projection
onto the feasible region $\Phi$. Consequently, the long-term variable
$\rho_{k}$ is updated as
\begin{equation}
\rho_{k}^{t+1}=(1-\beta_{t})\rho_{k}^{t}+\beta_{t}\bar{\rho}_{k}^{t}.\label{eq:update-rho}
\end{equation}
}
\emph{Remark 1} : Note that the stationary weight vector $v_{k}^{*}=\nabla_{\overline{\eta}_{k}}g_{k}\left(\tilde{\eta}_{k}^{\ast}\right)$
has captured the nature of the utility function. However, it is difficult
to obtain $\boldsymbol{v}^{*}$, since it in turn depends on the stationary
solution $\boldsymbol{\rho}^{*}$. Therefore, the basic idea of the
proposed algorithm is to iteratively update the long-term variable
$\boldsymbol{\rho}^{t}$ and the weight vector $\boldsymbol{v}^{t}$
such that $\boldsymbol{\rho}^{t}$ and $\boldsymbol{v}^{t}$ converge
to a stationary solution $\boldsymbol{\rho}^{*}$ and the corresponding
stationary weight vector $\boldsymbol{v}^{*}$, respectively.

\subsection{\textcolor{black}{Convergence Analysis}}

The following theorem states that Algorithm 2 converges to a stationary
point of $\mathcal{P}_{2}\left(\bm{\rho},\boldsymbol{v},\mathbf{\hat{H}}\right)$
up to certain convergence error which vanishes to zero exponentially
as $N\rightarrow\infty$.
\begin{thm}
[Convergence of Algorithm 2]\label{thm:Convergence-of-Short-term}
Suppose problem $\mathcal{P}_{2}\left(\bm{\rho},\boldsymbol{v},\mathbf{\hat{H}}\right)$
has a discrete set of stationary points, denoted by $\mathcal{F}^{*}\left(\bm{\rho},\boldsymbol{v},\mathbf{\hat{H}}\right)$.
Let $\mathbf{F}^{N}\left(\bm{\rho},\boldsymbol{v},\mathbf{\hat{H}}\right)$
denote the limiting point of the sequence generated by Algorithm 2
with input parameter $\bm{\rho},\boldsymbol{v},\mathbf{\hat{H}}$
and sample number $N$. Then for every small positive number $\epsilon>0$,
there exist positive constants $\hat{a}(\epsilon)$ and $\hat{b}(\epsilon)$,
independent of $N$, such that
\[
\mathrm{Pr}\left\{ \min_{\mathbf{F}\in\mathcal{F}^{*}\left(\bm{\rho},\boldsymbol{v},\mathbf{\hat{H}}\right)}\!\!\!\|\mathbf{F}^{N}\left(\bm{\rho},\boldsymbol{v},\mathbf{\hat{H}}\right)\!\!-\!\!\mathbf{F}\left(\bm{\rho},\boldsymbol{v},\mathbf{\hat{H}}\right)\!\|\!\geq\!\epsilon\!\right\} \!\leq\!p(\epsilon,N),
\]
for $N$ sufficiently large, where $p(\epsilon,N)\triangleq\hat{a}(\epsilon)e^{-N\hat{b}(\epsilon)}$.
\end{thm}
\begin{IEEEproof}
Specifically, the proposed FP-BCD algorithm falls in the MM framework
and similar proof is provided in \cite{Kaiming_arxiv19_D2D}. From
Theorem 4.4 in \cite{Jacobson_TIP07_MM}, every limiting point $\mathbf{F}^{N}$
of sequence generated by the short-term FP-BCD algorithm is a stationary
point of problem $\mathcal{P}_{3}\left(\bm{\rho},\boldsymbol{v},\mathbf{\hat{H}}\right)$,
where problem $\mathcal{P}_{3}\left(\bm{\rho},\boldsymbol{v},\mathbf{\hat{H}}\right)$
is the sample average approximation of problem $\mathcal{P}_{2}\left(\bm{\rho},\boldsymbol{v},\mathbf{\hat{H}}\right)$
with $N$ samples. As stated in \cite{Shapiro_SIAM09_SAA}, problem
$\mathcal{P}_{3}\left(\bm{\rho},\boldsymbol{v},\mathbf{\hat{H}}\right)$
is equivalent to problem $\mathcal{P}_{2}\left(\bm{\rho},\boldsymbol{v},\mathbf{\hat{H}}\right)$
w.p.1 when $N$ approaches to infinity, due to the classical law of
large number for random functions. That is to say, as $N\rightarrow\infty$,
any stationary point of $\mathcal{P}_{3}\left(\bm{\rho},\boldsymbol{v},\mathbf{\hat{H}}\right)$
is also a stationary point of problem $\mathcal{P}_{2}\left(\bm{\rho},\boldsymbol{v},\mathbf{\hat{H}}\right)$
w.r.1. When $N$ is finite, Algorithm 2 converges to approximate stationary
points of problem $\mathcal{P}_{2}\left(\bm{\rho},\boldsymbol{v},\mathbf{\hat{H}}\right)$
with the exponential convergence rate $\hat{a}(\epsilon)e^{-N\hat{b}(\epsilon)}$.
This is consequence of \cite{SAA_convergenceRate}, Theorem 3.1, which
provides a general convergence result for the original problem that
satisfies the following assumptions: (a) The feasible set of optimization
variables is a nonempty closed convex set; (b) The objective function
of the original problem is continuously differentiable on the feasible
set for any given random system states, and its gradient is Lipchitz
continuous. Clearly, problem $\mathcal{P}_{2}\left(\bm{\rho},\boldsymbol{v},\mathbf{\hat{H}}\right)$
satisfies the aforementioned assumption (a) and (b). This completes
the proof.
\end{IEEEproof}
Based on Theorem 2, the convergence of the proposed MO-SSCA algorithm
is summarized in the following theorem.
\begin{thm}
[Convergence of the Algorithm 1]\label{thm:Convergence-of-long}Given
problem (\ref{eq:originalP}), suppose that $\tau>0$ in (\ref{upsurrgate})
and the step-sizes $\{\alpha_{t}\}$ and $\{\beta_{t}\}$ are chosen
so that
\end{thm}
\begin{enumerate}
\item $\alpha_{t}\rightarrow0$, $\frac{1}{\alpha_{t}}\leq O\left(t^{\kappa}\right)$
for some $\kappa\in\left(0,1\right)$, $\sum_{t}\left(\alpha_{t}\right)^{2}<\infty,$
\item $\beta_{t}\rightarrow0$, $\sum_{t}\beta_{t}=\infty$, $\sum_{t}\left(\beta_{t}\right)^{2}<\infty$,
\item $\lim_{t\rightarrow\infty}\beta_{t}/\alpha_{t}=0$.
\end{enumerate}
Let $\{\bm{\rho}^{t},\boldsymbol{v}^{t},\mathbf{F}^{N}(i),\forall i\in\mathcal{T}_{t}\}_{t=1}^{\infty}$
denote the sequence of iterates generated by Algorithm 1, where $\mathbf{F}^{N}(i)\triangleq\mathbf{F}^{N}\left(\bm{\rho}^{t},\boldsymbol{v}^{t},\mathbf{\hat{H}}(i)\right),i\in\mathcal{T}_{t}$.
Then every limit point $\bm{v}^{*},\bm{\rho}^{*}$ of $\left\{ \bm{v}^{t},\bm{\rho}^{t}\right\} _{t=1}^{\infty}$
almost surely satisfies

\begin{equation}
\bm{v}^{\ast}=\nabla_{\overline{\bm{\eta}}}g\left(\overline{\bm{\eta}}^{\ast}\right),\label{eq:stationary-weight}
\end{equation}
\begin{equation}
(\boldsymbol{\rho}-\boldsymbol{\rho}^{*})^{T}\nabla_{\bm{\rho}}^{T}g\left(\bm{\overline{\eta}}\left(\bm{\rho}^{\ast},\bm{\Theta}^{N}\left(\bm{v}^{\ast},\bm{\rho}^{\ast}\right)\right)\right)\leq0,\forall\boldsymbol{\rho}\in\mathbf{\Phi},\label{eq:ooutconv}
\end{equation}
where $\bm{\Theta}^{N}\left(\bm{v}^{\ast},\bm{\rho}^{\ast}\right)\triangleq\left\{ \mathbf{F}^{N}(\bm{v}^{\ast},\bm{\rho}^{\ast},\mathbf{\hat{H}}),\forall\mathbf{\hat{H}}\right\} $,
and $\overline{\bm{\eta}}^{\ast}\triangleq\overline{\bm{\eta}}\left(\bm{\rho}^{\ast},\bm{\Theta}^{N}\left(\bm{v}^{\ast},\bm{\rho}^{\ast}\right)\right).$
Moreover, $\forall\mathbf{F}\in\varLambda\text{,}$ it satisfies
\begin{equation}
\left(\mathbf{F}\!-\!\mathbf{F}^{N}(i)\right)^{T}\!\!\!\!J_{\mathbf{F}}\left(\bm{\rho}^{\ast},\mathbf{F}^{N}(i)|\hat{\bm{H}}(i)\right)\!\nabla_{\overline{\bm{\eta}}}g\left(\overline{\bm{\eta}}^{\ast}\right)\!\leq e\left(N\right),\label{eq:inerconv}
\end{equation}
where $J_{\mathbf{F}}\left(\bm{\rho}^{\ast},\mathbf{F}^{N}(i)|\hat{\bm{H}}(i)\right)$
is the Jacobian matrix of the vector $\bm{\overline{\eta}}^{\circ}\triangleq[\overline{\eta}_{1}^{\circ},\cdots,\overline{\eta}_{K}^{\circ}]^{T}$
w.r.t. $\mathbf{F}$ at $\bm{\rho}=\bm{\rho}^{\ast}$ and $\mathbf{F}=\mathbf{F}^{N}(i)$,
and $e\left(N\right)$ satisfies $\lim_{N\rightarrow\infty}e\left(N\right)=0$
almost surely.
\begin{IEEEproof}
Based on Theorem 2, Theorem \ref{thm:Convergence-of-long} can be
proven by a similar approach in \cite{Anliu_tsp19_THCF}. Thus, we
omit the details due to the limited space.
\end{IEEEproof}
According to equation (\ref{eq:inerconv}) in Theorem 3, it implies
that the short-term solution $\mathbf{F}^{N}(i)$ found by Algorithm
2 must satisfy the stationary condition approximately with certain
error $e\left(N\right)$ that converges to zero exponentially as $N\rightarrow\infty$.
Moreover, the limiting point $(\bm{v}^{\text{\ensuremath{\ast}}},\bm{\rho}^{\ast})$
generated by Algorithm 1 also satisfies the stationary conditions
in (\ref{eq:stationary-weight}) and (\ref{eq:ooutconv}), respectively.
Thus, Algorithm 1 converges to stationary solutions of the mixed-timescale
optimization problem $\mathcal{P}$. Note that since $e\left(N\right)$
converges to zero exponentially, Algorithm 2 with a small $N$ can
already achieve a good performance and avoids excessive computational
complexity.

\section{Simulation Results and Discussions\label{sec:Simulation-Results-and}}

We consider a single-cell of radius $100$ m, where BS is equipped
with $64$ antennas. There are 12 devices randomly distributed in
the cell. We adopt a geometric channel model with a \textit{half-wavelength
space} ULA for simulations \cite{Anliu_tsp19_THCF}. The channel between
BS and device $k$ is given by $\boldsymbol{h}_{k}=\sum_{i=1}^{N_{p}}\varepsilon_{k,i}\boldsymbol{\textrm{a}}\left(\varphi_{k,i}\right)$,
where $\boldsymbol{\textrm{a}}\left(\varphi\right)$ is the array
response vector, $\varphi_{k,i}$'s are Laplacian distributed with
an angular spread $\sigma_{\textrm{AS}}=10$, $\varepsilon_{k,i}\sim\mathcal{CN}\left(0,\sigma_{k,i}^{2}\right)$,
$\sigma_{k,i}^{2}$ are randomly generated from an exponential distribution
and normalized such that $\sum_{i=1}^{N_{p}}\sigma_{k,i}^{2}=G_{n,k}$,
$G_{k}$ is the average channel gain determined by the pathloss model
$30.6+36.7\log10\left(\textrm{ds}_{k}\right)$ \cite{3gpp_Rel9},
and $\textrm{ds}_{k}$ is the distance between BS and device $k$
in meters. We consider $N_{p}=6$ channel paths for each device. The
transmit power budget for BS is $P_{\mathrm{max}}=10$ dBm. We set
$N=200$,\textcolor{black}{{} $S_{k}=24$ mW, $a_{k}=150$, $b_{k}=0.014$,
$\gamma_{k}=10$,} $\omega_{k}^{2}=-40$ dB, $\sigma_{k}^{2}=-60$
dBm and $\delta_{k}^{2}=-50$ dBm. There are $T_{s}=10$ time slots
in each frame and the slot size is 2 ms.\textcolor{black}{{} The coherence
interval $T=400$, which corresponds to a coherence time of 2 ms and
a coherence bandwidth of 200 kHz \cite{MLYN2016}.} The coherence
time for the channel statistics is assumed to be 10 s. We use the
average sum utility $g\left(\hat{\bm{\eta}}\right)=\sum_{k=1}^{K}\hat{\eta}_{k}$
as an example to illustrate the advantages of the proposed scheme.
Two schemes are included as baselines: 1) maximum ratio transmission
(MRT) scheme, which is obtained by fixing the MRT beamformer \cite{QJS_2014twc_JBPS};
2) zero-forcing (ZF) scheme, which is obtained by fixing the ZF beamformer
\cite{Dong_2018globecom_ZFSWIPT}. The power splitters of both MRT
and ZF scheme are obtained by the long-term optimization.

\begin{figure}[t]
	\centering
    \subfigure[]{
    \label{fig:varSNR}
    \includegraphics[height=2.5cm]{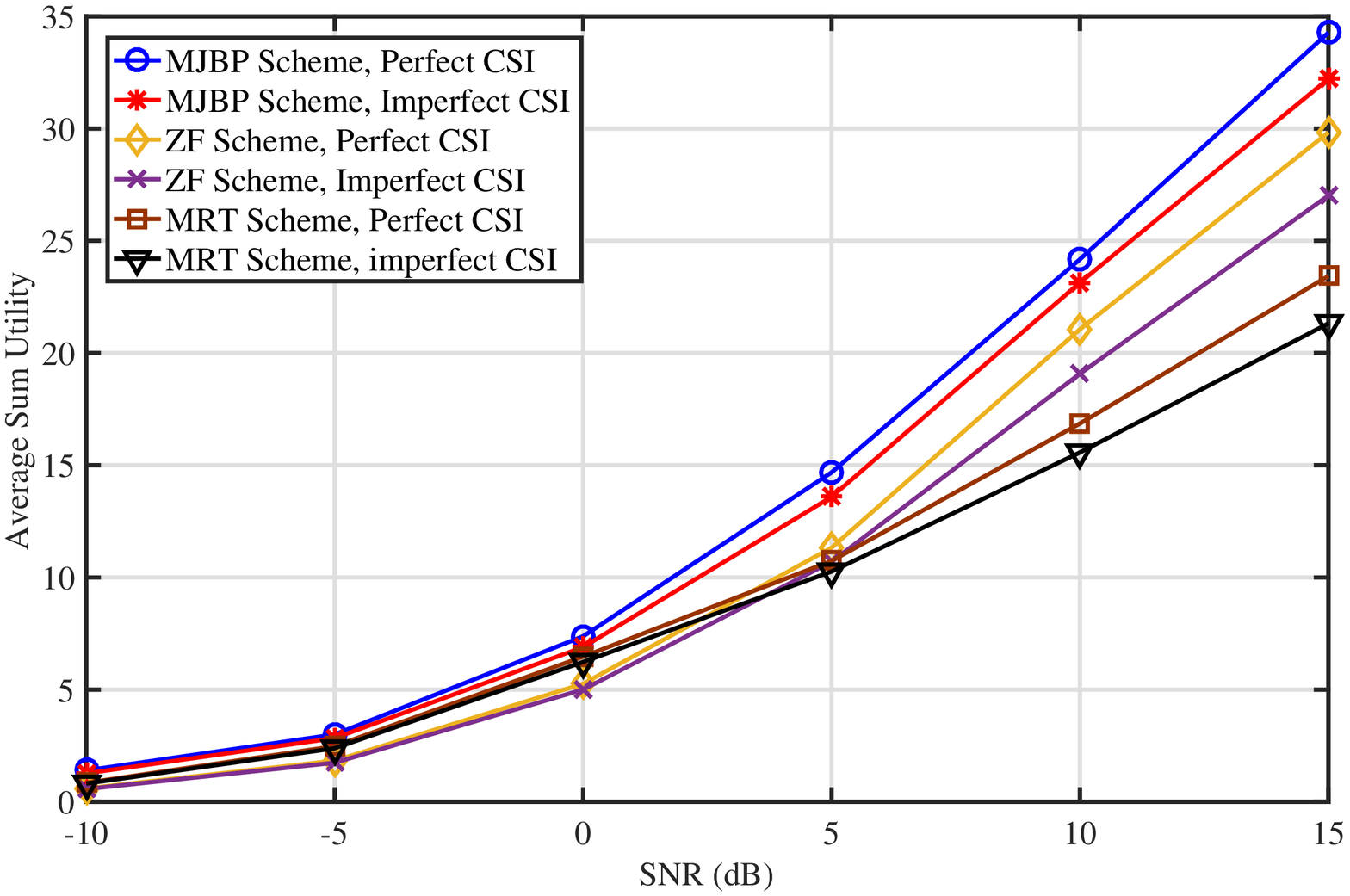}}
    \quad
    \subfigure[]{
    \label{fig:tradeoff}
    \includegraphics[height=2.6cm]{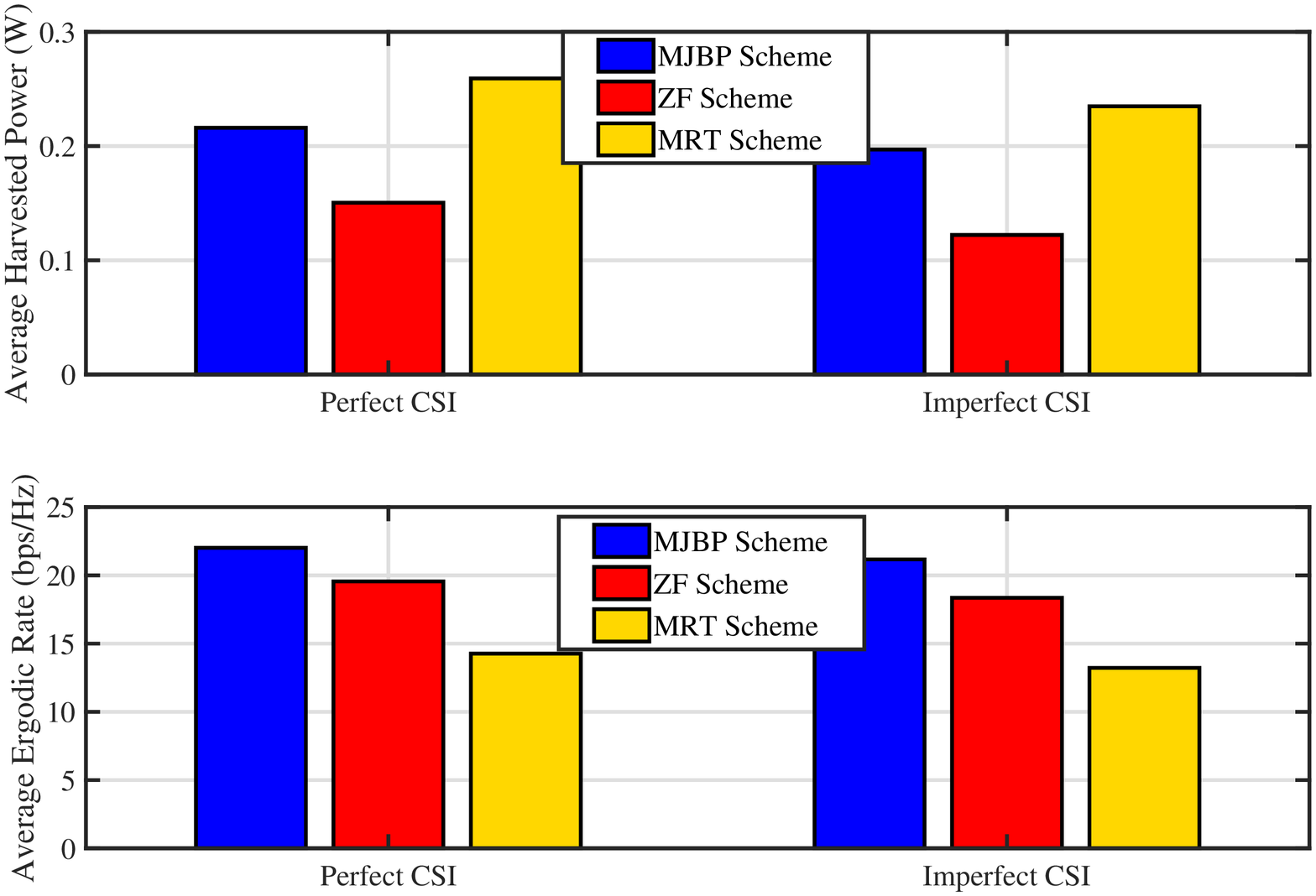}}
    \caption{(a) Utility performance versus SNR. (b) Tradeoff comparison for different schemes ($M=64$,$K=12$, and $\mathrm{SNR}=10$ dB).}
\end{figure}

In Fig \ref{fig:varSNR}, we plot the utility performance versus the
signal-to-noise ratio (SNR). We can see that as the SNR increases,
the average sum utility of all schemes increases gradually. It is
observed that the average sum utility achieved by the proposed MJBP
scheme is higher than that achieved by the other schemes for moderate
and large SNR. This indicates that the proposed MJBP scheme can better
mitigate the multi-device interference to achieve better tradeoff
between the average ergodic rate and the average harvested power,
\textcolor{black}{which is further validated in Fig \ref{fig:tradeoff}.}

\begin{figure}[t]
	\centering
    \subfigure[]{
    \label{fig:varK}
    \includegraphics[height=2.5cm]{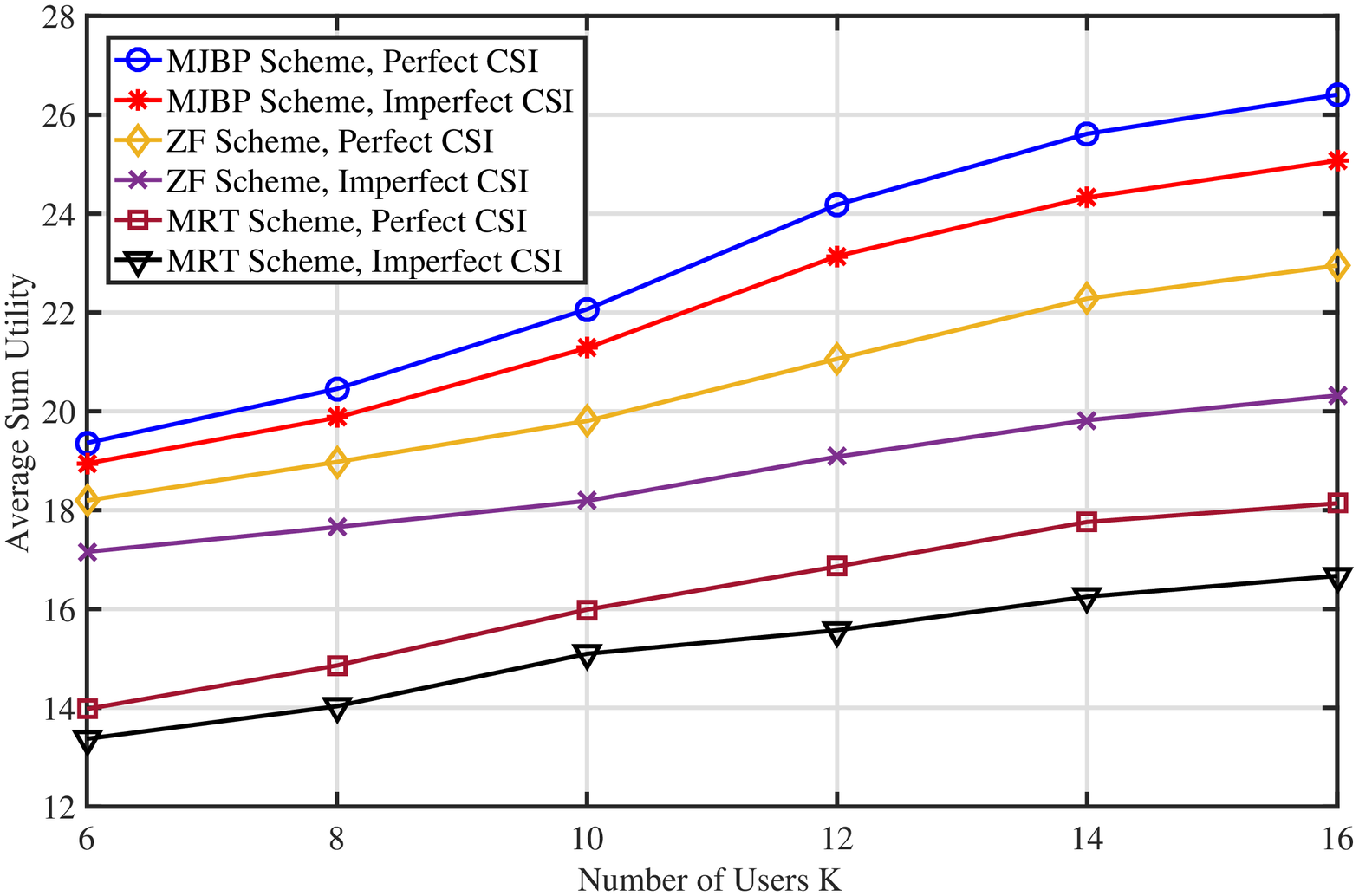}}
    \quad
    \subfigure[]{
    \label{fig:varM}
    \includegraphics[height=2.5cm]{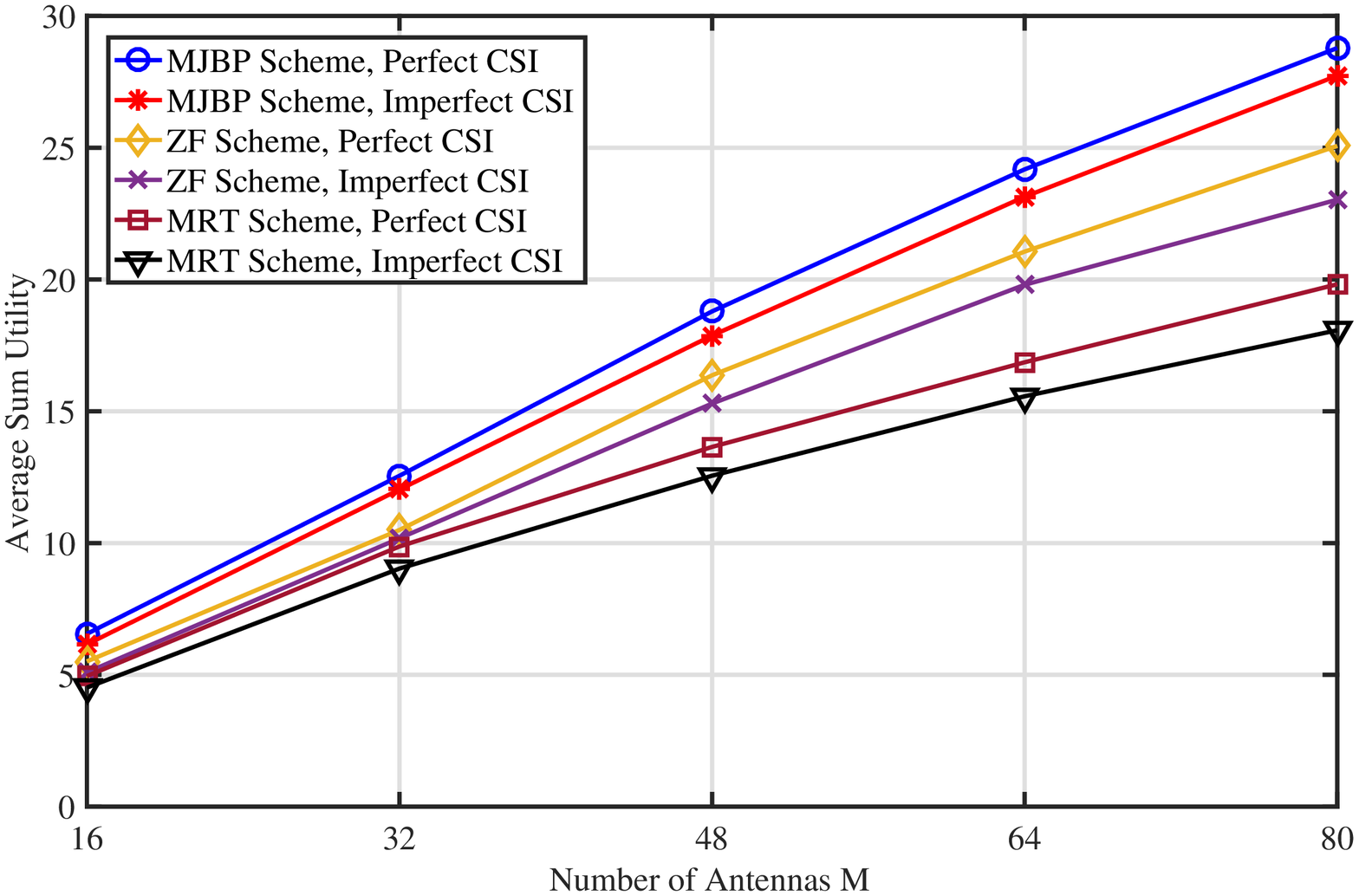}}
    \caption{(a) Utility performance versus the number of devices
$K$. (b) Utility performance versus the number of antennas
$M$.}
\end{figure}

In Fig \ref{fig:varK}, we plot the utility performance versus the
number of devices $K$. We observe that the proposed MJBP scheme achieves
significant gain over MRT scheme and ZF scheme, which demonstrates
the importance of mixed-timescale joint optimization. Moreover, as
the number of devices $K$ increases, the performance gap between
the proposed MJBP scheme and other competing schemes becomes larger.

\textcolor{black}{In Fig \ref{fig:varM}, we plot the utility performance
versus the number of antennas at BS. It shows that the performance
of all these schemes is monotonically increasing with the number of
antennas. Again, it is seen that the proposed MJBP scheme outperforms
all the other schemes for all $M$ regime.}

\section{Conclusion}

In this letter, we considered mixed-timescale joint beamforming and
power splitting (MJBP) scheme in the downlink transmission of massive
MIMO aided SWIPT IoT network to maximize the network utility under
the power budget constraint. We proposed a MO-SSCA algorithm to find
stationary solutions of the mixed-timescale non-convex stochastic
optimization problem. Simulations verify that the proposed MJBP scheme
achieves significant gain over existing schemes.

\bibliographystyle{IEEEtran}
\bibliography{liu_all}

% Generated by IEEEtran.bst, version: 1.13 (2008/09/30)
\begin{thebibliography}{10}
\providecommand{\url}[1]{#1}
\csname url@samestyle\endcsname
\providecommand{\newblock}{\relax}
\providecommand{\bibinfo}[2]{#2}
\providecommand{\BIBentrySTDinterwordspacing}{\spaceskip=0pt\relax}
\providecommand{\BIBentryALTinterwordstretchfactor}{4}
\providecommand{\BIBentryALTinterwordspacing}{\spaceskip=\fontdimen2\font plus
\BIBentryALTinterwordstretchfactor\fontdimen3\font minus
  \fontdimen4\font\relax}
\providecommand{\BIBforeignlanguage}[2]{{%
\expandafter\ifx\csname l@#1\endcsname\relax
\typeout{** WARNING: IEEEtran.bst: No hyphenation pattern has been}%
\typeout{** loaded for the language `#1'. Using the pattern for}%
\typeout{** the default language instead.}%
\else
\language=\csname l@#1\endcsname
\fi
#2}}
\providecommand{\BIBdecl}{\relax}
\BIBdecl

\bibitem{Swan_survey12_IoT}
M.~Swan, ``Sensor mania! the {Internet} of things, wearable computing,
  objective metrics, and the quantified self 2.0,'' \emph{J. Sens. Actuator
  Netw}, vol.~1, no.~3, pp. 217--253, 2012.

\bibitem{Ruizhang_2013twc_SpatialSWIPT}
R.~Zhang and C.~K. HO, ``M{IMO} broadcasting for simultaneous wireless
  information and power transfer,'' \emph{IEEE Trans. Wireless Commun.},
  vol.~12, no.~5, pp. 1989--2001, May 2013.

\bibitem{QJS_2014twc_JBPS}
Q.~Shi, L.~Liu, W.~Xu, and R.~Zhang, ``Joint transmit beamforming and receive
  power splitting for {MISO} {SWIPT} systems,'' \emph{IEEE Trans. Wireless
  Commun.}, vol.~13, no.~6, pp. 3269--3280, Apr 2014.

\bibitem{QJS_2014tsp_SOCP}
Q.~Shi, W.~Xu, T.~Chang, Y.~Wang, and E.~Song, ``Joint beamforming and power
  splitting for {MISO} interference channel with {SWIPT}: An {SOCP} relaxation
  and decentralized algorithm,'' \emph{IEEE Trans. Signal Process.}, vol.~62,
  no.~23, pp. 6194--6208, Dec 2014.

\bibitem{Dong_2018globecom_ZFSWIPT}
G.~Dong, H.~Zhang, and D.~Yuan, ``Optimal downlink transmission in {M}assive
  {MIMO} enabled {SWIPT} systems with zero-forcing precoding,'' \emph{IEEE
  Global Commun. Conf.}, Dec 2014.

\bibitem{Caire_2018twc_MIMObound}
G.~Caire, ``On the ergodic rate lower bounds with applications to masssive
  {MIMO},'' \emph{IEEE Trans. Wireless Commun.}, vol.~17, no.~5, pp.
  3258--3268, May 2018.

\bibitem{Anliu_tsp19_THCF}
A.~Liu, X.~Chen, W.~Yu, V.~Lau, and M.~Zhao, ``Two-timescale hybrid compression
  and forward for massive {MIMO} aided {C-RAN},'' \emph{IEEE Trans. Signal
  Process.}, vol.~67, no.~9, pp. 2484--2498, Mar 2019.

\bibitem{Ng_CL15_nonlinear}
E.~Boshkovska, D.~Ng, N.~Zlatanov, and R.~Schober, ``Practical non-linear
  energy harvesting model and resource allocation for {SWIPT} systems,''
  \emph{IEEE Commun. Lett.}, vol.~19, no.~12, pp. 2082--2085, Dec. 2015.

\bibitem{Shapiro_SIAM09_SAA}
A.~Shapiro, D.~Dentcheva, and A.~Ruszczynski, \emph{Lecture on stochastic
  programming: modeling and theory}.\hskip 1em plus 0.5em minus 0.4em\relax
  SIAM, 2009.

\bibitem{Kaiming_TSP18_FP}
K.~Shen and W.~Yu, ``Fractional programming for communication systems {--}
  {P}art {II}: {U}plink scheduling via matching,'' \emph{IEEE Trans. Signal
  Process.}, vol.~66, no.~10, pp. 2631--2644, Mar 2018.

\bibitem{Jacobson_TIP07_MM}
M.~W. Jacobson and J.~A. Fessler, ``An expanded theoretical treatment of
  iteration-dependent majorize-minimize algorithms,'' \emph{IEEE Transactions
  on Image Processing}, vol.~16, no.~10, pp. 2411--2422, Oct 2007.

\bibitem{CVX_MATLAB}
\BIBentryALTinterwordspacing
``{CVX} {R}esearch, {I}nc. {CVX:} {M}atlab software for disciplined convex
  programming, version 2.0 beta.'' Sep, 2012. [Online]. Available:
  \url{http://cvxr.com/cvx.}
\BIBentrySTDinterwordspacing

\bibitem{Kaiming_arxiv19_D2D}
K.~Shen, W.~Yu, L.~Zhao, and D.~P. Palomar, ``Optimal of {MIMO}
  device-to-device networks via matrix fractional programming: A
  minorization-maximization approach,'' \emph{arXiv.org:1808.05678}, 2019.

\bibitem{SAA_convergenceRate}
H.~Sun and H.~Xu, ``A note on uniform exponential convergence of sample average
  approximation of random functions,'' \emph{J. Math. Anal. Appl}, pp.
  698--708, 2012.

\bibitem{3gpp_Rel9}
\BIBentryALTinterwordspacing
\emph{Technical Specification Group Radio Access Network; Further Advancements
  for E-UTRA Physical Layer Aspects}, 3GPP TR 36.814. [Online]. Available:
  \url{http://www.3gpp.org}
\BIBentrySTDinterwordspacing

\bibitem{MLYN2016}
T.~Marzetta, E.~G. Larsson, H.~Yang, and H.~Ngo, \emph{Fundamentals of Massive
  {MIMO}}.\hskip 1em plus 0.5em minus 0.4em\relax Cambridge, U.K.: Cambridge
  University Press, 2016.

\end{thebibliography}

\end{document}